# Double-site Substitution of Ce into (Ba, Sr)MnO$_3$ Perovskites for Solar Thermochemical Hydrogen Production

Su Jeong Heo[1]*, Michael Sanders[2], Ryan O'Hayre[2], and Andriy Zakutayev[1]*

[1]Materials Science Center, National Renewable Energy Laboratory, Golden, CO 80401 USA

[2]Metallurgical and Materials Engineering, Colorado School of Mines, Golden, CO 80401 USA

**Corresponding Author**

*Su Jeong Heo: SuJeong.Heo@nrel.gov, Andriy Zakutayev: Andriy.Zakutayev@nrel.gov

**ABSTRACT**

Solar thermochemical hydrogen production (STCH) is a renewable alternative to hydrogen produced using fossil fuels. While serial bulk experimental methods can accurately measure STCH performance, screening chemically complex materials systems for new promising candidates is more challenging. Here we identify double-site Ce-substituted (Ba,Sr)MnO$_3$ oxide perovskites as promising STCH candidates using a combination of bulk synthesis and high-throughput thin film experiments. The Ce substitution on the B-site in 10H-BaMnO$_3$ and on the A-site in 4P-SrMnO$_3$ lead to 2-3x higher hydrogen production compared to CeO$_2$, but these bulk single-site substituted perovskites suffer from incomplete reoxidation. Double-site Ce substitution on both A- and B-site in (Ba,Sr)MnO$_3$ thin films increases Ce solubility and extends the stability of 10H and 4P structures, which is promising for their thermochemical reversibility. This study demonstrates a high-throughput experimental method for screening complex oxide materials for STCH applications.

**TOC GRAPHICS**

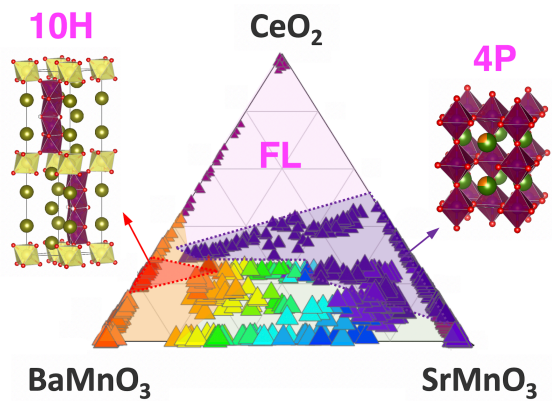

Hydrogen is a valuable chemical precursor and storable energy carrier that is widely used in ammonia production and fuel cell vehicles.[1,2] Hydrogen produced by water electrolysis using renewable electricity is a sustainable alternative to steam reforming of natural gas.[3-7] Among several emerging hydrogen production routes (e.g., photoelectrochemical, electrolysis), solar thermochemical hydrogen production (STCH) has received considerable attention.[8-10] One of the possible STCH cycles is a two-step water-splitting process, involving partial thermal reduction of a suitable metal oxide, followed by steam based reoxidation for hydrogen production. Non-stoichiometric pure and doped ceria ($CeO_2$) compounds have been widely investigated for this STCH process.[11-13] However, the low extent of ceria reduction required hydrogen production under very high temperatures (>1500 °C), increasing the difficulty of engineering suitable reactors.[14]

Oxide materials with perovskite and perovskite-related structures are promising alternative redox candidates for replacing ceria in STCH applications.[15,16] The $ABO_3$ perovskite composition can accommodate a variety of A- and B-site cations and structure variations, enabling broad opportunities to control the extent of oxygen reduction.[17,18] Perovskite oxides tend to easily form point defects such as oxygen vacancies, which facilitates fast redox reaction kinetics due to higher oxygen exchange capacities, and can also enable an increased extent of reduction at lower temperatures.[19] Among currently known redox-active perovskites, $BaCe_{0.25}Mn_{0.75}O_{3-\delta}$ (BCM25)[20], $SrTi_{0.5}Mn_{0.5}O_{3-\delta}$ (STM55)[21], $LaGa_{0.4}Co_{0.6}O_{3-\delta}$ (LGC60)[22], $La_{1-x}Sr_xMnO_3$ (LSM)[23,24], and $Sr_xLa_{1-x}Mn_yAl_{1-y}O_{3-\delta}$ (SLMA)[25] have shown particular promise for STCH application.

The staggering number of oxide perovskites compositions motivates the need for high-throughput research methods to discover the most promising perovskites for STCH application, and probe composition-structure-property relations that can improve understanding of the STCH process. Numerous studies have sought new STCH perovskites using high-throughput computational methods.[26,27] However, experimental (combinatorial) high-throughput method have been less explored for perovskites in STCH compared to other applications (e.g., piezoelectrics[28]). In addition to the combinatorial screening for STCH application studied here, the oxide perovskite crystal structure and redox properties show great promise for a wide range of other applications including fuel cells,[29] catalysis,[30] and electrochemical water splitting.[31]

Here, we investigate double-site Ce substitution of (Ba,Sr)MnO$_3$ perovskites for solar thermochemical hydrogen production using a combination of serial bulk synthesis and high-throughput thin film methods. Temperature programed reduction and water splitting experiments with single-site Ce substituted 10H-BaMnO$_3$ and 4P-SrMnO$_3$ bulk powder samples, suggest that double-site Ce substitution of (Ba,Sr)MnO$_3$ may increase Ce incorporation and improve the redox stability of these materials for STCH applications. Guided by a Goldschmidt tolerance factor model, thin film combinatorial experiments with double-site substituted (Ba,Sr)$_{1-x}$Ce$_x$Ce$_y$Mn$_{1-y}$O$_3$ are used to evaluate composition-structure-processing trends in this complex 4-cation chemical space. We find that co-doping Ce on both A-site and B-site in (Ba,Sr)MnO$_3$ increases overall Ce solubility and increases composition stability range of 10H and 4P structures, both of which should lead to higher STCH performance. This study provides new understanding of composition-structure-processing-property relations in the (Ba,Sr,Ce)(Mn,Ce)O$_3$ material system, and demonstrates an efficient way to uncover new STCH-active compounds using high-throughput experimental methods.

***Bulk powders***

Single-site Ce-substituted $BaMnO_3$ and $SrMnO_3$ oxide perovskites were synthesized as bulk powders, including $BaCe_{0.25}Mn_{0.75}O_{3-\delta}$ (BCM25) and $Sr_{0.75}Ce_{0.25}MnO_{3-\delta}$ (SCM25) using a sol-gel modified Pechini method[32] (see supporting information for experimental details). The resulting crystal structures and the relative phase fractions were determined from X-ray diffraction (XRD) measurements and Rietveld refinement, as shown in Figure 1a. This quantification procedure revealed that the SCM25 bulk powder has almost phase-pure 4P structure (99.0%) while the BCM25 bulk powder is predominantly 10H polymorph with 7% of 12R and 7% of fluorite (FL) secondary phases. This 10H polymorph of BCM25 is likely to convert to the 12R polymorph after several redox cycles as shown in prior investigations.[20]

The redox potential of these bulk compositions was compared using temperature-programmed reduction (TPR), with results presented in Figure 1b. The behavior of the single-site substituted BCM25 and SCM25 are compared to $SrMnO_3$ (SMO) simple perovskite and ceria ($CeO_2$), which is the current benchmark materials for STCH. While undoped ceria has a too low extent of reduction ($\Delta\delta < 0.03$) but fully reoxidizes, undoped SMO has exceptionally large extents of reduction but the driving force for reoxidation is probably too low to split water. Ce-doped $BaMnO_3$ and $SrMnO_3$ starts to reduce at considerably lower temperature (< 950 °C), with large extents of reduction ($\Delta\delta > 0.17$ for BCM25 and $\Delta\delta > 0.26$ for SCM25) compared to ceria which is beneficial for STCH. However, both BCM25 and SCM25 do not fully reoxidize ($\Delta\delta > 0.01$) compared to ceria, which is a disadvantage for STCH applications.

Water-splitting hydrogen production of single-site substituted BCM25 and SCM25 perovskites was measured under two reduction/oxidation condition sets using a stagnation flow reactor (SFR) that has been described elsewhere.[33] The results are compared to undoped ceria baseline material and double-site substituted SLMA perovskites measured at the same conditions,[20] as shown in Figure 1c. BCM25 releases a larger amount of $H_2$ at $T_{RE}$ 1350 °C and $T_{OX}$ 850°C than SCM25, and SCM25 produces a larger amount of hydrogen at $T_{RE}$ 1400°C and $T_{OX}$ 1050 °C than BCM25.

The amounts of $H_2$ produced by both BCM25 and SCM25 perovskite compounds are 2-3x larger than by undoped ceria. However, the $H_2$ production of these single-site substituted perovskites is lower than double-site La and Al co-substituted $SrMnO_3$ (SLMA), which is one of the most promising perovskites for STCH.[34,35] Similarly to double-site substitution in SLMA, the STCH performance in $Ba_{1-x}Sr_xCo_yFe_{1-y}O_{3-\delta}$ [36] and $Ca_{0.5}Ce_{0.5}MO_3$ (M = 3d transition metals) [37] material systems has been suggested and demonstrated to increase STCH performance. Thus, it is likely that similar improvements of STCH performance should be possible in the double site substituted $(Ba,Sr)_{1-x}Ce_xCe_yMn_{1-y}O_3$ perovskite in the compositional stability range at high temperature.

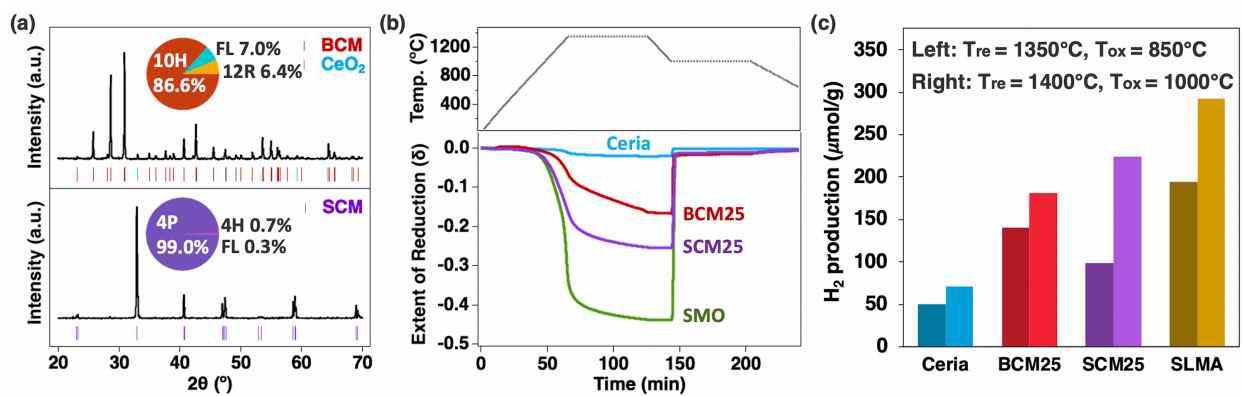

**Figure 1**. Characterization of STCH-active perovskites and ceria in bulk powder form: (a) XRD and Rietveld refinement for the BCM25 and SCM25 powders used as end-member targets, (b) Temperature Programmed Reduction for single-site substituted BCM25 and SCM25 and undoped $CeO_2$ and $BaMnO_3$

bulk samples, and (c) STCH performance of SCM25 powders and other compounds reported elsewhere[20] for the conditions of $T_{red}$=1350°C & $T_{ox}$=850°C and $T_{red}$=1400°C & $T_{ox}$=1000°.

***Substitution model***

To predict the structural trends in the (Ba,Sr,Ce)(Ce,Mn)O$_3$ perovskite materials system studied here, we use the Goldschmidt tolerance factor. The Goldschmidt tolerance factor[38] reflects the degree of the ideal cubic perovskite distortion that can arise from a size-mismatch of cations and anions, defined as $t = \frac{r_A + r_O}{\sqrt{2}\,(r_B + r_O)}$ where $r_A, r_B$, and $r_O$ represent the average ionic radii of the A, B, and O-sites, respectively. Generally, cubic symmetry with minor distortion of the corner-sharing BO$_6$ octahedra is found for compounds where the tolerance factor lies between roughly 0.9 and 1.0. As the value of $t$ increases above 1, the ratio of face-sharing to corner-sharing octahedra increases to accommodate the correspondingly larger A-site cations and/or smaller B-site cations which can lead to the formation of hexagonal structures.

Figure 2 illustrates the modeled changes in tolerance factor within the (Ba,Sr,Ce)(Ce,Mn)O$_3$ chemical composition quaternary (Fig. 2a) and pseudo-ternary (Fig. 2b) phase diagrams containing the BaMn$_{1-y}$Ce$_y$O$_3$ (BCM), Sr$_{1-x}$Ce$_x$MnO$_3$ (SCM), Ba$_{1-x}$Sr$_x$MnO$_3$ (BSM), and (Ba,Sr)$_{1-x}$Ce$_x$Mn$_{1-y}$Ce$_y$O$_3$ (BSCMC) compositional families of interest to this study. The BaMnO$_3$ has a calculated Goldschmidt tolerance factor ($t$) of 1.103 which is far greater than unity, so BaMnO$_3$ forms a hexagonal 2H structure.[39] With the B-site substitution of smaller Mn ions by larger Ce ions in BaMnO$_3$, the 1.103 tolerance factor gradually decreases and reaches 1.056 at 25% Ce incorporation in Ba(Ce,Mn)O$_3$ resulting in the 10H structure[40] (BCM25). The tolerance factor of BaMnO$_3$ is also reduced with A-site substitution of smaller Ba ions by smaller Sr ions along the BaMnO$_3$/SrMnO$_3$ tie-line, ending with t = 1.041 for pure strontium manganate's 4H structure.[41]

To further decrease the tolerance factor, it is possible to reduce some of the $Mn^{4+}$ to $Mn^{3+}$ by partially substituting the $Sr^{2+}$ by $Ce^{3+}$ on A-site of $SrMnO_3$ while keeping the charge balance for stability. In this case, the simultaneous increase of the B-site and decrease of the A-site mean ionic radii lead to rapid lowering of the tolerance factor, reaching 0.993 for $Sr_{0.5}Ce_{0.5}MnO_3$, which shows a tetragonal-like perovskite structure (*I4/mcm*).[42] The tolerance factor can be also lowered by increasing the compositional complexity of the material with simultaneous A and B-site doping. As shown in Figure 2, starting from BCM25, the tolerance factor is gradually decreased by substituting Sr on A-site while it more steeply decreases by substituting both Sr and Ce in $(Ba,Sr)(Ce,Mn)O_3$. Finally, combining all of these effects, the $(Ba,Sr,Ce)(Ce,Mn)O_3$ compositions is of interest in this study to achieve smaller tolerance factors by moving from BCM25 towards higher concentrations of Sr and Ce.

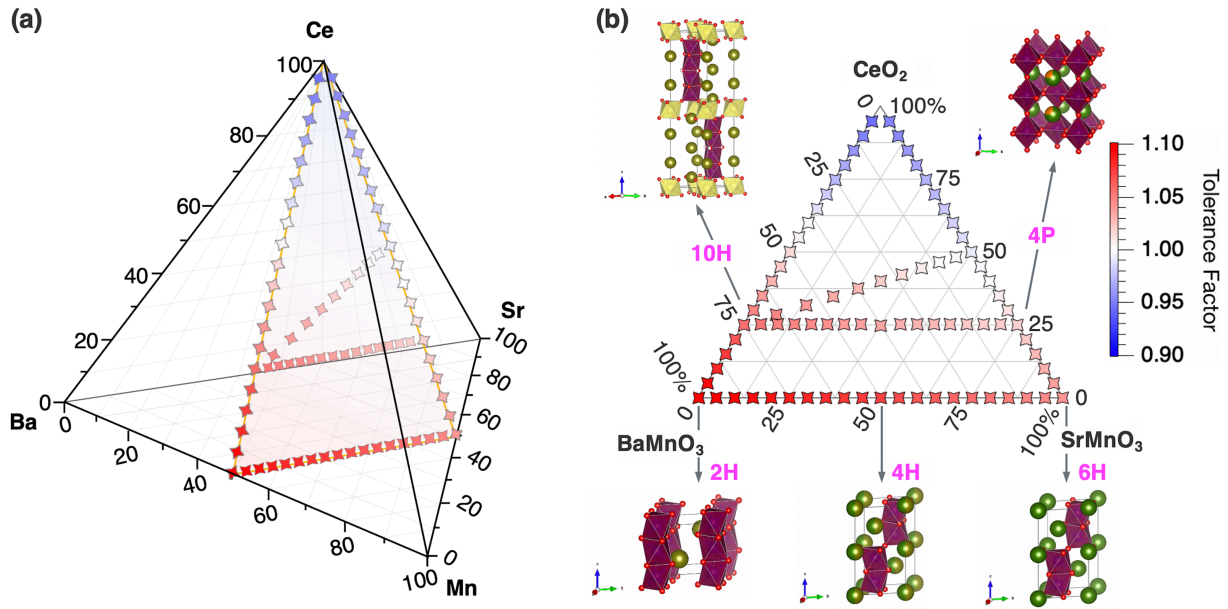

**Figure 2**. Model of single- and double-site Ce substitution into $(Ba,Sr)MnO_3$ perovskites: (a) Goldschmidt tolerance factor over the composition space of the BSCM quaternary phase diagram and (b) $BaMnO_3$-$SrMnO_3$-$CeO_2$ ternary phase diagram with corresponding crystal structures. Single- and double-site Ce

substitution of (Ba,Sr)MnO$_3$ may induce a structural change due to reductions in the tolerance factor. CeO$_2$ has a fluorite structure (FL), so the perovskite tolerance factor values have decreasing significance as the compositions shift towards pure ceria.

### *Single site substitution*

To validate the predictions of our (Ba,Sr,Ce)(Ce,Mn)O$_3$ structural model (Fig. 2), we used combinatorial pulsed laser deposition (PLD) to synthesize sample libraries with many chemical compositions (Figure 3),[43] and anneal them at elevated temperature in air. These sample libraries are evaluated for their composition, structure and properties using spatially-resolved measurement techniques (see supporting information for experimental details). Figure 3a shows the pseudo-ternary phase diagrams of the structure of BaCe$_y$Mn$_{1-y}$O$_3$, Sr$_{1-x}$Ce$_x$MnO$_3$, and Ba$_{1-x}$Sr$_x$MnO$_3$ pseudo-binary tie-lines along its edges. The structures of these perovskite-related oxides can be distinguished by X-ray diffraction in the range of 2θ = 30.7-33.2º based on small peak shifts (Figure 3b). For example, in the case of BaMnO$_3$ in Figure 3a, a two-layer hexagonal (2H) structure featuring face-shared oxygen polyhedron with a space group of *P*6$_3$/*mmc*,[39] has been determined from its signature XRD peak at 31.4 degrees.

As shown in Figure 3a, the 2H-BaMnO$_3$ B-site substitution of Mn$^{4+}$ by Ce$^{4+}$ or Ce$^{3+}$ along the BaCe$_y$Mn$_{1-y}$O$_3$ tie-line leads to the formation of a 10H-BaCe$_{0.25}$Mn$_{0.75}$O$_3$ (10H-BCM25) structure reported to be promising for STCH[20], which is formed via face-sharing octahedral [Mn$_4$O$_{15}$] tetramers that share corners with [Ce$_{0.83}$Mn$_{0.17}$O$_6$] octahedral along the *c*-axis.[40] The 2H-BaMnO$_3$ A-site replacement of larger Ba$^{2+}$ (*r* = 1.61 Å in 12-fold coordination) with smaller Sr$^{2+}$ (r = 1.44 Å in 12-fold coordination) leads to the 4H-polymorph of SrMnO$_3$ (Figure 3a), which contains both corner- and face-sharing octahedra.[44-46] In between BaMnO$_3$ and SrMnO$_3$, the Sr$_x$Ba$_{1-x}$MnO$_3$ alloy, with the varying oxygen stoichiometry determined by the ratio of corner-sharing (cubic) to face-sharing (hexagonal) octahedra, may offer additional STCH candidates with tailored oxygen vacancy formation energies. Substituting the A-site $Sr^{2+}$ in 4H-$SrMnO_3$ with smaller $Ce^{3+}$ ($r$ = 1.34 Å, CN=12) and/or $Ce^{4+}$ ($r$ = 1.14 Å, CN=12) leads to a structural phase transition from 4H hexagonal symmetry ($P6_3/mmc$) to 4P tetragonal-like symmetry ($I4/mcm$).[47]

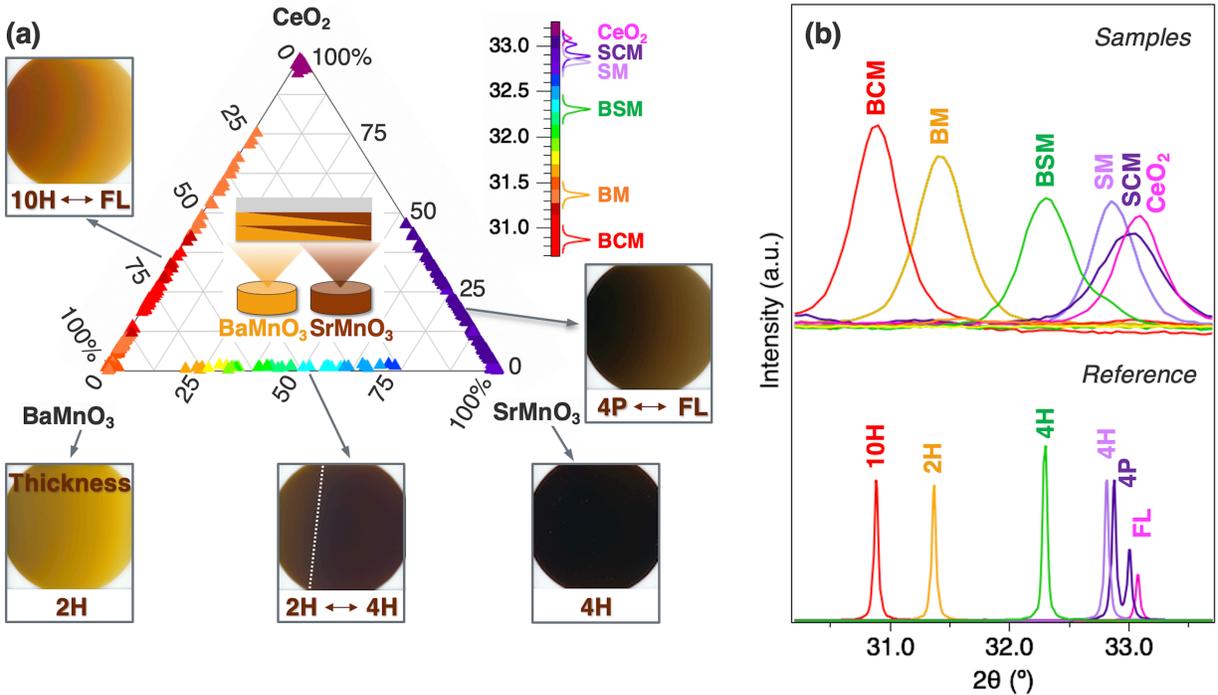

**Figure 3**. The (Ba,Sr,Ce)(Ce,Mn)$O_3$ thin-film pseudo-ternary phase diagram: (a) XRD 2θ position of the primary peaks mapped over the composition space of the $BaMnO_3$-$SrMnO_3$-$CeO_2$ ternary phase diagram with corresponding film images. (b) Representative sample XRD patterns of the primary peaks of each sample and reference patterns for comparison. The individual structures in this complicated phase space can be distinguished by XRD in the range of 2θ = 30.7-33.2°.

In addition to XRD, the color has been measured for the compositionally graded thin films annealed at 950 °C (Fig. 3a) using an optical scanner,[48] which is an important metric because it potentially can be used to evaluate the extent of reduction in thermochemical cycling. The pale-green color of 2H-BaMnO$_3$ thin films is consistent with the dark-green color of bulk powder.[39] As Ce is substituted on B-site of 2H-BaMnO$_3$, the film color changes to bright-orange when it forms the 10H structure at the BCM25 composition, and changes to a light-yellow color at higher Ce-fractions. The dark-brown color of 4H-SrMnO$_3$ is in good agreement with a bulk powder report,[49] and also lightens as Ce is substituted on A-site of 4H-SrMnO$_3$ to form the 4P structure of SCM25 with inclusions of FL at yet-higher Ce fraction. As the annealing temperature increases above 950 °C, the overall film colors became darker, followed by separation on black and white as expected from segregation of the CeO$_2$ fluorite (FL) secondary phases.

### *Double site substitution*

Now that the site substitution model for (Ba,Sr)$_{1-x}$Ce$_x$Ce$_y$Mn$_{1-y}$O$_3$ has been established (Fig. 2) and validated (Fig. 3), we explore the potential to further increase the Ce concentration in the oxide perovskite materials by substituting the A-site with Ba in (Sr,Ce)MnO$_3$ and with Sr in Ba(Ce,Mn)O$_3$, which should enhance the stability of the perovskite structure and lead to improved STCH performance. The pseudo-ternary (Ba,Sr)$_{1-x}$Ce$_x$Ce$_y$Mn$_{1-y}$O$_3$ phase diagram in Figure 4 shows the dominant structure determined from XRD 2θ position (color scale map) and relative perovskite phase intensity-ratio (marker size) within the BaMnO$_3$-SrMnO$_3$-CeO$_2$ phase space after annealing at elevated temperatures.

Along the BaMnO$_3$-CeO$_2$ tie line in Figure 4a, the 10H structure is observed in a wide Ce-concentration range (about 13-42%) at 950 °C, but with low perovskite phase ratio above 25% Ce. When Sr is incorporated in Ba(Ce,Mn)O$_3$, the 10H structure region increases to higher Ce concentration (~32%) with higher perovskite phase ratio. Since BCM is reported to form a line compound with an exact Ce/Mn ratio of 0.25/0.75 due to size constraints on the B-site[50], this result shows that excess Ce above 25% may be accommodated by substitution on the A-site. With increasing temperature, the Sr-free 10H structure (red color in Fig. 4) diminishes and eventually disappears at 1150 ºC (in Fig. 4c). However, the Sr-containing 10H structure transitions to the 4P structure in this high temperature range. This suggests that Sr-substituted Ba(Ce,Mn)$O_3$ with 4P structure is a new promising candidate for STCH applications, in accordance with our double site substitution model (Figure 2)

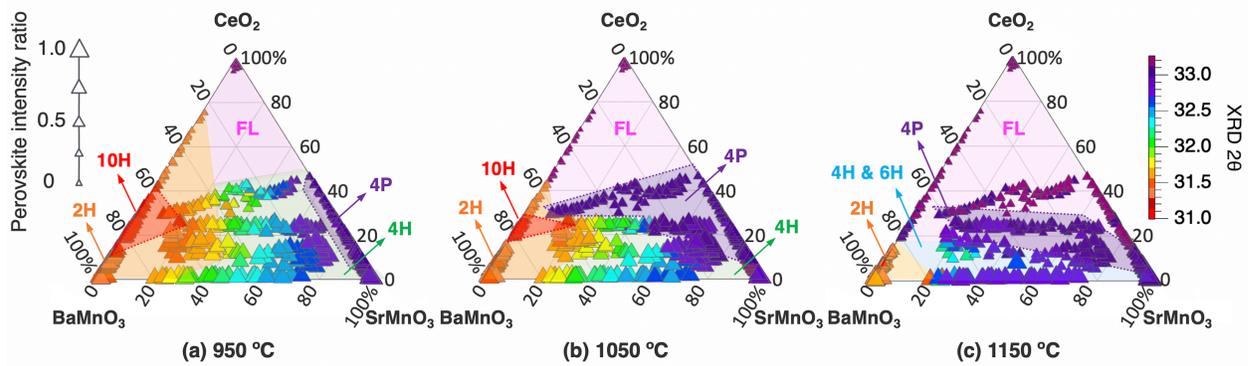

**Figure 4**. Double-site Ce substitution into thin-film (Ba,Sr)MnO$_3$ perovskites, showing XRD 2θ position (color scale map) and perovskite phases-ratio (marker size) measured over the composition space of BM-SM-Ce oxide ternary phase diagram after 950 ºC (a), 1050 ºC (b), and 1150 ºC (c) annealing in air. Double site Ce substitution into (Ba,Sr)MnO$_3$ increases the composition range of stability for 10H and 4P structures suitable for STCH as illustrated for bulk powders in Fig.1.

Along the SrMnO$_3$-CeO$_2$ tie line in Figure 4, the 4P structure appears when Ce is substituted in (Sr,Ce)MnO$_3$ and persists up to 55% Ce-concentration at 950 ºC (Fig. 4a) and 1050 ºC (Fig. 4b), while the perovskite-ratio decreases at higher Ce-concentration and at the highest temperature of 1150 ºC (Fig. 4c). As discussed above, the 4P structure can accommodate very high Ba concentration (max. ~92% at 1050 ºC) into (Sr, Ce)MnO$_3$ which can be observed in the center area of the phase diagram in Figure 4. Although the 4P structure coexists with FL structure, the FL-ratio decreases upon Ba-doping at the same Ce-concentration line (horizontal lines in Fig. 4), implying that the Ba substitution permits additional Ce to be accommodated into the 4P structure. Again, this suggests that Ba-substituted (Sr,Ce)MnO$_3$ with 4P structure may be promising for STCH applications due to a broader compositional range of stability and its ability to accommodate higher Ce content compared to BCM.

### *Conclusions*

In summary, the screening of the (Ba, Sr, Ce)(Mn, Ce)O$_3$ space by a combination of serial bulk synthesis and high-throughput thin film experimentations indicates that the composition stability range of tetragonal 4P-polymorph of (Sr,Ce)MnO$_3$ can be significantly enhanced by up to 90% Ba-substitution even at temperatures >1000 °C. In addition, the compositional stability range of the hexagonal 10H-polymorph is enhanced by substitution of Sr into Ba(Ce,Mn)O$_3$, while the total Ce fraction is increased, and a FL secondary phase content is reduced. These thin film results are supported by bulk powder measurement of high thermochemical water splitting performance of these 4P (e.g. SCM25) and 10H (e.g. BCM25) which is 2-3x larger compared to the ceria baseline material. Our work not only demonstrates an experimental high-throughput approach to identify new redox materials for renewable hydrogen production, but also provides general insights into composition-structure-property relations for designing oxide perovskites by double site substitution for other energy applications.

ASSOCIATED CONTENT

**Supporting Information**. Bulk experimental methods, including powder synthesis and their XRD, Rietveld refinement, TPR characterization, and water splitting measurements. Thin film experimental methods including combinatorial deposition of single- and double-site substituted perovskite by PLD system with post annealing, spatially-resolved XRD and XRF for mapping of the perovskite films.

AUTHOR INFORMATION

Corresponding Authors

Su Jeong Heo - Materials Science Center, National Renewable Energy Laboratory, Golden, CO 80401 USA; orcid.org/0000-0002-7933-9714; Email: SuJeong.Heo@nrel.gov

Andriy Zakutayev - Materials Science Center, National Renewable Energy Laboratory, Golden, CO 80401 USA; orcid.org/0000-0002-3054-5525; Email: Andriy.Zakutayev@nrel.gov

Authors

Michael Sanders - Metallurgical and Materials Engineering, Colorado School of Mines, Golden, CO 80401 USA; orcid.org/0000-0001-6366-5219; Email: misander@mymail.mines.edu

Ryan O'Hayre - Metallurgical and Materials Engineering, Colorado School of Mines, Golden, CO 80401 USA; orcid.org/0000-0003-3762-3052; Email: rohayre@mines.edu

**Notes**

The authors declare no competing financial interest.

ACKNOWLEDGMENT

This work was authored in part by the National Renewable Energy Laboratory (NREL), operated by Alliance for Sustainable Energy LLC, for the U.S. Department of Energy (DOE) under contract no. DE-AC36-08GO28308. Funding provided by the Office of Energy Efficiency and Renewable Energy (EERE) Hydrogen and Fuel Cell Technologies Office (HFTO), as a part of HydroGEN Energy Materials Network (EMN) consortium. The authors would like to acknowledge the use of a stagnation flow reactor (SFR) at Sandia National Laboratories, which is a multimission laboratory managed and operated by National Technology and Engineering Solutions of Sandia, LLC, a wholly owned subsidiary of Honeywell International Inc., for the U.S. Department of Energy's National Nuclear Security Administration under contract DE-NA0003525. We would also like to thank Dr. Yun Xu for preliminary data and initial discussions related to this work. The views expressed in the article do not necessarily represent the views of the DOE or the U.S. Government.